\documentclass[aps,prd,reprint,article,floatfix,nofootinbib]{revtex4-1}
\pdfoutput=1
\usepackage[usenames]{color} %used for font color
\usepackage{amssymb} %maths
\usepackage{amsmath} %maths
\usepackage{slashed}
\usepackage[utf8]{inputenc} %useful to type directly diacritic characters
\usepackage{simpler-wick}
\usepackage{physics}
\usepackage{graphicx}
\usepackage{subcaption}
\usepackage{fancyhdr}
\begin{document}
\centering
\begin{flushleft}
\title{Minimum length universe in a Robertson Walker geometry}
\author{Dragana Pilipovi\'{c}}
\affiliation{Department of Physics, University of Illinois Chicago}
\date{\today}
\begin{abstract}
A minimum length universe is enforced via a diffusive Markovian field. Requiring the invariance of the proper time functional under spacetime transformations generated by such fields provides us with many new terms added to the Christoffel Connection, Equations of Motion, and Friedmann Equation. Under the Robertson-Walker (RW) metric, we obtain species-specific energy-density evolution, now a function of both the parameters defining the diffusive fields as well as the RW scale factor. \\
\text{}\\
The math leads us to a relationship between temporal diffusion, universe expansion acceleration, and the speed of universe expansion, along with a derivation of the cosmological constant as a function of both minimum length and the scalar field expectation value of the Higgs mechanism.\\
\text{}\\
The theory gives us a means of relating the Hubble parameter to its diffusive ML 'equivalent', and a means to express the effective equation of state in terms of the deceleration parameter, q.  We find that q can take on a range of higher values without invoking the need for a dark matter species, and that the universe inhabits a vacuum or matter dominated regime as a function of q.\\
\end{abstract}
\maketitle
\textbf{{I. INTRODUCTION}}\\
\text{}\\
Robertson Walker Minimum Length (RWML) theory corresponding to a negative deceleration parameter, $q$, and a positive Hubble parameter, $H$, requires Markovian ML fields to transform at a different scale to that of spacetime.  While both spacetime and ML fields are, independently, conformally Minkowski in a local frame ($r=0$), relative to each other the two conformal spaces scale at a non-zero power of the RW scale factor.  In simultaneously incorporating the effects of RW geometry and ML, the scale factor can no longer be conformally 'transformed away'.\\
\text{}\\
RWML theory opens up some new possibilities and introduces some new concepts.  It allows for a greater possible range of the deceleration parameter and the effective universe equation of state, with corresponding matter vs. vacuum dominated regimes.  RWML also introduces a new diffusion parameter $D$, the diffusion 'equivalent' of $H$, and a potentially new species, the Higgs vacuum.  \\
\text{}\\
Specifically, we set out to answer the following questions:\\
\text{}\\
a) why is the vacuum energy density so small relative to what one might expect given large scalar field expectation values in Quantum Field Theory (QFT)?\\
\text{}\\
b) what is the origin of the small positive energy density of the vacuum?\\
\text{}\\
c) how can we explain the relative ratios of energy density parameters across species, specifically why is $\rho_M/\rho_V \sim 0.5$ while $\rho_R/\rho_M \sim 10^{-3}$ [1]?\\
\text{}\\
\color{black}d) is it possible to have a deceleration parameter, $q$, which is greater in magnitude than allowed by Robertson-Walker (RW) theory ($q=-\frac{1}{2}$) without requiring a new (dark matter) species with $w<-1$?\color{black}\\
\text{}\\
\text{}\\
RWML suggests the following answers:\\
\text{}\\
a) ML temporal diffusion is the driver of both the universe expansion and its acceleration.  But also, at the minimum possible vacuum energy, the expansion velocity is directly proportional to the QFT scalar field expectation value, $\bar{v}$. ML theory in a RW geometry thus gives us the relationship between temporal diffusion, $\bar{v}$, and the acceleration of universe expansion.  The resulting RW scale factor is at present extremely large, giving us a very small vacuum energy.\\
\text{}\\
b) ML spatial diffusion gives the vacuum energy density a minimum positive value proportional to $\bar{v}^2$.  The RWML theory further shows us that the vacuum energy density is an irreducible quadratic function of $H$ and $D$ over the reals, with $H > 0$, but only if the ML fields transform at a different power of the scale factor relative to that of spacetime.  At this minimum possible positive vacuum energy density, the velocity of expansion is positive and defined by spatial diffusion, and hence $\bar{v}$.\\
\text{}\\
c) Given the present energy density ratios across the species, we can estimate that the magnitude of the rate of spatial diffusion for radiation is roughly three times that of the rate of diffusion for matter.\\
\text{}\\
\color{black}d) Under RWML theory we find that - without invoking any new species - the universe has matter-dominated and vacuum-dominated regimes as a function of the acceleration parameter $f, f\equiv -q$.  We find an open\footnote{While we work in a local, and thus geometrically flat spacetime, under RWML the universe can still be open or closed as a relative function of $D$ and $H$.} and mostly vacuum-dominated universe for $f$ in the neighborhood of one, with an effective equation of state described by $w \approx - 0.7$. Lower values for $f$ put the universe in the approximately constant $D$ regime with an exactly matter-dominated effective equation of state, $w=0$, for the choice of $f=\frac{1}{2}$. \\
\text{}\\ 
RWML further suggests that were the acceleration parameter to grow well beyond one, universe has a choice of one of two paths.  One path allows for a 'peaceful' transition between matter and vacuum dominated states as long as $H$ remains positive.  The second path is a universe 'dooms-day' scenario providing a degenerate set of solutions at higher values of the acceleration parameter but for $H<0$.\\
\text{ }\\
\text{ }\\
\textbf{{II. MOTIVATION FOR RWML UNIVERSE}}\\  
\text{}\\
While RW theory without ML diffusion gives us a curious static solution to the Friedmann equations by requiring that $w = - \frac{1}{3}$, this is not a problem in an expanding universe.  However, the ad hoc addition of the cosmological constant in the case of a mostly flat ($\kappa \approx 0$) universe makes the RW theory incomplete.  On the other hand, ML diffusion under a Minkowski metric ($\eta$) gives us no means by which we can relate the metric expansion and its acceleration to the ML diffusive process.  Each theory - on its own - lacks the exact characteristics of the other.  \\
\text{}\\
ML diffusion under an RW metric ($g$)  describes an universe naturally characterized by minimum positive energy under a flat geometry (we work in $txyz$ coordinates, a local frame $r = 0$, and a mostly positive RW metric $g_{\mu \nu} = (-1, a^2, a^2, a^2)$):
\begin{align*}
& (R^{ML}_{\mu \nu})_{\eta} = (0, 9 c^2 \sigma^4, 9 c^2 \sigma^4, 9 c^2 \sigma^4) \rightarrow (R^{RWML}_{\mu \nu})_{g} \tag{1}\\
& c^2 \sigma^4 = \frac{\pi g_\sigma^4 \bar{v}^2}{72} \ \approx \ 55 \ GeV^{2} \ \tag{2} 
\end{align*}
The spatial diffusion parameters, $c \sigma^2$, take on the values defined by the Higgs mechanism in QFT, where $g_\sigma \approx 0.38$ is the rate of diffusion corresponding to the Higgs field,\footnote{Also the third generation of fermions and the QCD regime.  See D. Pilipovi\'{c}, et al. forthcoming papers on QFT/Higgs aspects of ML theory.} and $\bar{v} \approx 246 \  GeV$ is the scalar field expectation value.  \\
\text{}\\
By working in the RW geometry, we are able to relate the curvature created by ML diffusion, and captured by the spatial diffusion parameter $D\equiv \frac{c \sigma^2}{a^2}$, to the Hubble parameter $H \equiv \frac{\dot{a}}{a}$, as well as define a relationship between the effective equation of state parameter $w$ and the acceleration parameter.\\
\text{}\\ 
\text{}\\
\textbf{{III. POSITIVE EXPANSION ACCELERATION }}\\
\text{}\\
The resulting RWML universe exhibits an expansion acceleration driven by temporal diffusion parameter $c_\epsilon$, with the following relationship to the RW scale factor $a_t$ in the case of a vacuum dominated universe ($w=-1$):
\begin{align*}
& \frac{3}{2} c_\epsilon = \frac{\ddot a}{a} + (\frac{\dot{a}}{a})^2 = (f+1) H^2,\\
& H \equiv \frac{\dot{a}}{a}, f \equiv (\frac{\ddot a}{a})^{-1} H^2 \tag{3}
\end{align*}
Note that if we can approximately express the acceleration term in (3) using the acceleration factor $f$, the above gives us an exponential scale factor.  For an acceleration parameter close to $\frac{1}{2}$, hence $\frac{\ddot a}{a} \approx \frac{1}{2}(\frac{\dot a}{a})^2$, we have $a_t \sim e^{\sqrt{c_\epsilon} t}$.\\
\text{}\\
For a positive temporal diffusion parameter, equation (3) puts a lower bound on the acceleration parameter, $f>-1$.  However, entirely based on (3), there appears to be no upper bound on $f$: even for a constant $c_\epsilon$, the increase in $f$ can be directly offset with a decrease in $H^2 \sim \frac{1}{f+1}$. \\
\text{}\\
As is shown in later section 4. (see equation (11)), at the point of minimum energy density for an approximately constant $D$, we obtain a relationship between the Hubble parameter $H$ and the spatial diffusion parameters $c \sigma^2$, giving us further a relationship between temporal diffusion, spatial diffusion, and the acceleration of the universe expansion:
\begin{align*}
& H|_{\rho_{min}} = \frac{3}{4} \frac{c \sigma^2}{a^2} \ \ \ \Rightarrow \ \ \ \frac{3}{2} c_\epsilon  = \frac{\ddot a}{a} + \frac{9}{16} \frac{c^2 \sigma^4}{a^4}  \tag{4} \\
&(\frac{\ddot a}{a})_0 \approx \frac{1}{2} H_0^2 \ \ \ \Rightarrow \ \ \ ({c_\epsilon})_0 \approx H_0^2   \tag{5} 
\end{align*}
The approximate present value for the temporal diffusion parameter in (5) above is based on the often favored approximation for the acceleration parameter, $f \approx \frac{1}{2}$.  We can thus establish the scale of the temporal diffusion parameter:
\begin{align*}
&c_\epsilon \sim 10^{-84} \ GeV^2 \tag{6} 
\end{align*}
The introduction of an RW metric under ML is critical in establishing the relative scales between the temporal and spatial diffusion parameters: while $c \sigma^2 \sim 10 \ GeV$, $\sqrt{c_\epsilon} \sim 10^{-42} \ GeV $.  These scales do not changes as we allow for higher values of the acceleration parameter $f$, or varying universe regimes.\\
\text{}\\
Furthermore, ML under a Minkowski metric gives us a torsionless universe with zero temporal diffusion, but ML under RW metric gives us an approximately torsionless universe given the extremely small temporal diffusion ($R_{0i} \sim c_\epsilon \sigma_\epsilon^2 H$).\\
\text{}\\
\textbf{{IV. SPECIES' ENERGY DENSITY RATIOS}}\\
\text{}\\
In addition to giving us the scale factor and temporal diffusion relationship in (3) for a vacuum dominated universe, equations of motion, $\nabla_\mu T^{\mu \nu} = 0$, also formulate the evolution of species' energy density.  For a universe with a roughly flat geometry, $\kappa << a^2$, we have:
\begin{align*}
& \frac{\dot{\rho}_w}{{\rho}_w} = - 3 \{(1+w) \frac{\dot{a}}{a} + \frac{c_w \sigma_w^2}{a^2} (2 + 3w) \}\tag{7} 
\end{align*}
A static solution to (7) corresponding to the Hubble parameter value in (4) defines the new Higgs-vacuum, with an equation of state given by $w=-\frac{11}{15} \approx -0.7333$. The additional possible static solution remains the physical vacuum devoid of diffusion, with $c_V \sigma_V^2 = 0$ and $w=-1$.  \\
\text{}\\
If we further allow matter to take on the same value for diffusion rate as the Higgs, $(g_\sigma)_M \approx g_\sigma$, roughly imposing species energy density ratios $\rho_M/\rho_V \sim 0.5$ and $\rho_R/\rho_M \sim 10^{-3}$, we find that the magnitude of diffusion rate for radiation is roughly three times that of matter: $\frac{|(g_\sigma)_R|}{(g_\sigma)_M}\approx 3.1$, given that initial density is uniform across species..\\
\text{}\\
In contrast to a universe with two static solutions, consider the case where $\frac{\dot{a}}{a} << \frac{c_w \sigma_w^2}{a^2}$.  Now a specie's energy density evolution is dominated by the species-specific spatial rates of diffusion, $\frac{c_w \sigma_w^2}{a^2}$, which are further functions of the specie's rates of diffusion ($c_w \sigma_w^2 \sim (g_\sigma)_w^2$).  In this case,\footnote{As is explained in Section 6.4, the spatial diffusion parameter $\sigma$ is a function of the scale factor $a_t$, and hence should be brought inside the time integral.  For rough back-of-the-envelope calculations here, we keep things simple instead.} 
\begin{align*}
& \rho_V = (\rho_V)_0 \ e^{3 c_V \sigma_V^2 \int \frac{dt}{a^2}}  \tag{8} \\
&\frac{\rho_M}{\rho_V} \approx (\frac{\rho_M}{\rho_V})_0 \ e^{- 3 (2 c_M \sigma_M^2 +  c_V \sigma_V^2)\int \frac{dt}{a^2}} \approx \frac{1}{2}  \tag{9} \\
&\frac{\rho_R}{\rho_M} \approx (\frac{\rho_R}{\rho_M})_0 \ e^{- 3 (3 c_R \sigma_R^2 -  2 c_M \sigma_M^2)\int \frac{dt}{a^2}} \approx 10^{-3}  \tag{10} 
\end{align*}
By setting the rate of diffusion for matter to roughly equal that of a vacuum, we can estimate $\int \frac{dt}{a^2} \approx \frac{ln 2}{9 \sqrt{55}} \approx 0.01$, given that initial density is uniform across species.  In this case also, the magnitude of the radiation rate of diffusion appears to be roughly three times the rate of diffusion for matter: $\frac{|(g_\sigma)_R|}{(g_\sigma)_M} \approx 3.3$.\\
\text{}\\
\text{}\\
\textbf{{V. UNIVERSE UNDER A SLOW MOVING $D$}}\\
\text{}\\
For an approximately constant diffusion parameter $D$, we minimize the energy density relative to $H$.  This scenario corresponds to $f \lessapprox 0.7998$, hence it includes the typically considered case of $f\approx \frac{1}{2}$.  In the next section we will consider the higher-valued acceleration parameters which allow us to properly minimize energy density to obtain a deeper understanding of how universe regimes relate to the acceleration parameter.\\
\text{}\\
For an RW metric with a zero curvature parameter $\kappa$, hence a flat RW geometry\footnote{A zero-curvature RW geometry corresponds to a conformally Minkowski space and a geometrically flat universe.} in accordance with both a vacuum-dominated universe and the current experimental observations, the cosmological constant acquires a minimum positive value defined by $D^2$ and directly related to the QFT (Higgs mechanism) scalar field expectation value:
\begin{align*}
& \frac{8 \pi G}{3}\rho = (H - \frac{3}{4} D)^2 + \frac{63}{16} D^2 \tag{11} 
\end{align*}
where $\rho$ is the total energy density.  For vacuum, matter, and radiation being the only physical species, at the present values based on observations, we have $\rho = \rho_V + \rho_M + \rho_R \approx \frac{3}{2} \rho_V$, giving us the present value for the cosmological constant:
\begin{align*}
& \Lambda = \frac{63}{8} \frac{c^2 \sigma^4}{a^4} = \frac{7 \pi}{64} \frac{ g_\sigma^4 \bar{v}^2}{a^4} \tag{12} 
\end{align*}
Thus, for slow moving $D$, the universe has a small vacuum energy density with a positive Hubble parameter,
\begin{align*}
& H = \frac{\dot{a}}{a} = \frac{3}{4} \frac{c \sigma^2}{a^2} > 0 \tag{13}
\end{align*}
\text{}\\
By applying the Friedmann Equations, we can solve for the equation of state parameter $w$ as a function of $f$, and thus obtain the effective species for the chosen acceleration parameter under the particular theory in question.  In case of the RW theory, with a choice of $f = \frac{1}{2}$, we find that $w=-\frac{2}{3}$ is the effective species, corresponding to expected mostly vacuum and matter mix.  It is interesting that the equation of state for the ML Higgs vacuum matches the RW effective universe equation of state.\\
\text{}\\
In contrast, RWML theory with an approximately constant $D$ and with $f=\frac{1}{2}$ gives us $w=0$ as the solution.  In this case, RWML Friedmann Equations describe a matter-dominated universe. \\
\text{}\\
Clearly the above approach is flawed if we cannot approximate the diffusion parameter as constant.  In fact, for $f \gtrapprox 0.7998$ not only can we not assume that $D$ is approximately constant, but the minimum energy density state becomes well defined, albeit degenerate.\\
\text{}\\
\text{}\\
\textbf{{VI. ACCELERATION AND UNIVERSE REGIMES}}\\
\text{}\\
On two separate occasions above we have taken a simplified approach to the diffusion parameters by setting them constant.  However, the two differential equations (3) and (13), as well as the requirement on how spatial diffusion parameter $\sigma$ must transform from Minkowski to RW geometry (see section VII.D.), tell us that this cannot be so, and that the diffusion parameters must be functions of time.  Equipped with the knowledge of what's to come (see section VI.), the ratio of the spatial diffusion parameters relative to the temporal diffusion parameter grows with time:\\
\begin{align*}
& c_\epsilon \sim \frac{(f+1)}{a^{12}}, \ \ \ {c \sigma^2} \sim \frac{1}{a^{4}}, \ \ \ \frac{c \sigma^2}{\sqrt{c_\epsilon}} \sim \frac{a^{2}}{ \sqrt{f+1}} \tag{14} 
\end{align*}
\text{}\\
Both $H$ and $D$ are functions of time.  In fact, since $\dot{H} = -(1-f) H^2$ and $D \sim \frac{1}{a^6}, \dot{D} =  - 6 H D$, we see that in the neighborhood of $f = 1$ it is $H$ which can be assumed relatively constant.  By minimizing the energy density, $\dot{\rho} = 0$, with the additional help of RWML Friedmann equations, we obtain both the relationship between the diffusive and Hubble parameters, $D$ and $H$, as well as the relationship between the acceleration parameter and the equation of state parameter, $f$ and $w$.  We find that for the acceleration parameter in the neighborhood of one, we obtain a universe with an effective $w \approx - 0.7$ and thus well approximated by the RW theory.\\
\text{}\\
Allowing the acceleration parameter to grow under RWML provides us with two different paths for $D$ (and thus also $H$): the first path corresponds to an effective equation of state with $w\approx -0.7$ in the neighborhood of $f = 1$.  This path also allows for a smooth transition from a mostly vacuum dominated universe to matter dominated, with a bounded energy density described by $H=\frac{3}{4} D$, as long as the Hubble parameter remains positive.  \\
\text{}\\
The second path gives us an unacceptable answer for $f=1$ by requiring that $D=0$: $D$ is an inverse function of $a$, and as such can get extremely small, but never zero.  As acceleration parameter grows, this second path describes a dooms-day scenario with a negative Hubble parameter and an infinitely growing energy density.\\
\text{}\\
Finally, we must consider at what point the universe might transition from open to closed.  While we are assuming a geometrically flat universe by working in local coordinates ($r=0$), the presence of ML fields introduces a curvature which allows the universe to be open or closed despite the geometric flatness.  We find that the turning point occurs at $D = \frac{H}{3}$ with a corresponding acceleration parameter $f=3$: for lesser values of the diffusion parameter or acceleration we have an open universe and otherwise closed.\\
\text{}\\
\textbf{{VII. SUPPORTING DERIVATIONS}}\\  
\textbf{{{A. Markovian fields in Minkowski geometry}}}\\  
\text{}\\
We enforce ML scale by introducing stochastic behavior in spacetime.  Each point in spacetime, $x$, under minimum length scale obtains a stochastic term: $x^\mu \rightarrow x^\mu + \sigma \delta \xi_x^\mu$.  The Markovian field $\delta \xi_x^\mu$ represents the uncertainty at $x$ due to minimum length scale.  It scales with the square root of the minimum length scale: $\delta \xi^\mu \sim \sqrt{\delta t}$, consistent with the behavior of stochastic variables. \\
\text{}\\
The second moment of such a field thus scales with $\delta t$, and is unique to each point in spacetime: 
\begin{align*}
&\langle \delta \xi_x^\alpha \ \delta \xi_y^\beta \rangle = \delta^4(x-y) \ g^{\alpha \beta} \ \delta t \tag{15} 
\end{align*}
The derivative of a stochastic field is not a 'typical' derivative at all, but rather a difference of two 'new' Markovian fields which we will refer to as the 'kinetic Markovian fields'.  For example, for a partial derivative in the direction of $\delta x^\mu$ we have
\begin{align*}
&\partial_{\mu} \delta \xi_x^\alpha = \frac{\delta \xi_{x+\delta x^\mu/2}^\alpha - \delta \xi_{x-\delta x^\mu/2}^\alpha}{\delta x^\mu} = \\
& \ \ \ \ (\delta \zeta_{x+\delta x/2})^\alpha_\mu - (\delta \zeta_{x-\delta x/2})^\alpha_\mu \tag{16} 
\end{align*}
The 'new' field $\delta \zeta_{x}$ is also a Markovian field, but one which scales with the square root of the inverse of minimum length scale, $(\delta \zeta_x)^\alpha_\mu \sim \frac{1}{\sqrt{\delta x^\mu}}$.  Its second moment now follows:
\begin{align*}
&\langle (\delta  \zeta_x)_\mu^\alpha \ (\delta  \zeta_y)_\nu^\beta \rangle |_{\mu = \nu} = \delta^4(x-y) \ g^{\alpha \beta} \ \frac{1}{\delta x^\mu},  \\
& \ \ \ \ \ \ \  \langle (\delta  \zeta_x)_\mu^\alpha \ (\delta  \zeta_y)_\nu^\beta \rangle |_{\mu \neq \nu} = 0 \tag{17} 
\end{align*} 
\text{}\\
\textbf{\emph{{B. ML connection}}}\\  
\text{}\\
Regardless of the metric of choice or the theory of choice, the single universal law, $\mathcal{U}^2 = -1$, defines the action $I$\footnote{See [3], pg. 106-107 for a detailed derivation of the proper time functional used here.} for a free particle moving along a world line. We introduce Markovian fluctuations into the classical theory of General Relativity by allowing for a Markovian (stochastic) translation, with the requirement that $I$ must be conserved: 
\begin{align*} 
& I = - \frac{1}{2} \int d\tau \frac{d x^\mu}{d \tau} g_{\mu \nu} \frac{d x^\nu}{d \tau}, \\
& \ \ \ \ \ \ \ \ x^\mu \rightarrow x^\mu + \delta \overline{x}^\mu + \sigma \delta {\xi}^\mu, \ \ \ \delta \xi^\mu \sim \mathcal{N}(0,\sqrt{\delta t}) \tag{18}\\
& I \rightarrow - \frac{1}{2} \int d\tau \{ \frac{d(x^\mu + \delta \overline{x}^\mu + \sigma \delta \tilde{\xi}^\mu)}{d \tau} \\
& \ \ \ \ \ \ \ \ \ \ \ \ \ \ \ \ \ \ (g_{\mu \nu} + (\partial_\rho g_{\mu \nu}) (\delta \overline{x}^\rho + \sigma \delta \tilde{\xi}^\rho)  \\
& \ \ \ +\frac{1}{2} (\partial_\alpha \partial_\beta g_{\mu \nu}) \sigma^2 \mathnormal{h}^{\alpha \beta}_\rho \delta \overline{x}^\rho )\frac{d(x^\nu + \delta \overline{x}^\nu + \sigma \delta \tilde{\xi}^\nu)}{d \tau} \} \tag{19}\\
& \langle \delta I \rangle = \int d\tau g_{\mu \rho} \delta \overline{x}^\rho \{ \frac{d^2 x^\mu}{d \tau^2} + \tilde{\Gamma}^{\mu}_{\alpha \beta} \frac{d x^\alpha}{d \tau} \frac{d x^\beta}{d \tau} \} = 0 \tag{20}\\
& \tilde{\Gamma}^{\mu}_{\alpha \beta} = {\Gamma}^{\mu}_{\alpha \beta} - \mathcal{H}^{\mu}_{\alpha \beta} \tag{21}
\end{align*}
The new Christoffel connection now includes the 'standard' Christoffel connection defined entirely by the metric, but also new terms due to ML fluctuations:
\begin{align*}
& \Gamma^{\mu}_{\alpha \beta} = \frac{1}{2} g^{\mu \sigma} (\partial_\alpha g_{\beta \sigma} + \partial_\beta g_{\sigma \alpha} - \partial_\sigma g_{\alpha \beta }) \tag{22}\\
& \mathcal{H}^{\mu}_{\alpha \beta} = \frac{1}{2} g^{\mu \delta} \{ \frac{1}{2} \sigma^2  \mathnormal{h}^{\eta \sigma}_{\delta} (\partial_\eta \partial_\sigma g_{\alpha \beta}) \\
& \ \ \ \ \ +  (\partial_\alpha \sigma^2) (\partial_\sigma g_{\eta \beta}) \mathnormal{h}^{\eta \sigma}_{\delta} + (\partial_\alpha \sigma) (\partial_\beta \sigma) \mathnormal{h}^{\eta \sigma}_{\delta} g_{\eta \sigma} \\
& \ \ \ \ \ +2 \sigma^2 a^{\sigma \eta}_{\delta \alpha} (\partial_\sigma g_{\eta \beta}) +  (\partial_\alpha \sigma^2) a^{\eta \sigma }_{\delta \beta} g_{\eta \sigma }\\
& \ \ \ \ \  +   \sigma^2 \mathnormal{k}^{\eta \sigma}_{\delta \alpha \beta} g_{\eta \sigma } \} \tag{23}\\
& \kappa^{\mu}_{\alpha \beta} \equiv \frac{1}{2} g^{\mu \delta} \sigma^2 \mathnormal{k}^{\eta \sigma}_{\delta \alpha \beta} g_{\eta \sigma } \tag{24}\\
& \mathcal{A}^{\mu}_{\alpha \beta} \equiv \frac{1}{2} g^{\mu \delta} \sigma^2 a^{\sigma \eta}_{\delta \alpha} (\partial_\sigma g_{\eta \beta})\tag{25}\\
&\Sigma ^{\mu}_{\alpha \beta} \equiv \frac{1}{4} g^{\mu \delta} \sigma^2  \mathnormal{h}^{\eta \sigma}_{\delta} (\partial_\eta \partial_\sigma g_{\alpha \beta}) \tag{26}
\end{align*}
where $h,k,a$ are tensors corresponding to various two-point functions described in detail in 6.3. and 6.4. below.  For an isotropic vacuum diffusion parameters must be constant in space: $\sigma(x) = \sigma$, but are not required to be constant in time.  In fact, as we shall see, the spatial diffusion parameter $\sigma$ is in fact a function of time, with $\sigma \sim \frac{1}{a^2}$.  However, we also find that all terms with $\partial_t \sigma$ are either $O(\sigma_\epsilon^4)$ or $O(DH^2)$ and can be approximated away.\footnote{See the Appendix for detail on these terms.}  \\
\text{}\\
\text{}\\
\textbf{\emph{C. ML Under Minkowski Metric}}\\
\text{}\\
For ML in a Minkowski geometry, the problem simplifies further:
\begin{align*}
& g_{\mu \nu} = \eta _{\mu \nu} \Rightarrow {\Gamma}^{\mu}_{\alpha \beta} = 0, \mathcal{H}^{\mu}_{\alpha \beta} = \kappa^{\mu}_{\alpha \beta}, \tilde{\Gamma}^{\mu}_{\alpha \beta} = - \kappa^{\mu}_{\alpha \beta} \tag{27}
\end{align*}
Equations (22)-(26) above assume a tensorial formulation of the various necessary two-point functions for the ML fields, as follows:
\begin{align*}
& \langle \delta {\xi}^\mu  \delta {\xi}^\nu \rangle \equiv  \mathnormal{h}^{\mu \nu}_{\rho} \delta \overline{x}^\rho, \ \ \ \  \langle \partial_\alpha \delta {\xi}^\mu  \partial_\beta \delta {\xi}^\nu \rangle \equiv \mathnormal{k}^{\mu \nu}_{\rho \alpha \beta} \delta \overline{x}^\rho , \\
& \ \ \ \ \ \ \ \ \langle \delta {\xi}^\mu  \partial_\alpha \delta \tilde{\xi}^\nu \rangle \equiv  \mathnormal{a}^{\mu \nu}_{\rho \alpha} \delta \overline{x}^\rho \tag{28}\\ 
& \langle \frac{d \delta {\xi}^\mu}{d \tau}  \frac{d \delta {\xi}^\nu }{d \tau}\rangle = \langle \partial_\alpha \delta {\xi}^\mu  \partial_\beta \delta {\xi}^\nu \rangle \frac{d x^\alpha}{d \tau} \frac{d x^\beta}{d \tau} =  \\
& \ \ \ \ \ \ \ \ \mathnormal{k}^{\mu \nu}_{\rho \alpha \beta} \frac{d x^\alpha}{d \tau} \frac{d x^\beta}{d \tau} \delta \overline{x}^\rho, \tag{29}\\
&\langle \delta {\xi}^\mu  \frac{d \delta \tilde{\xi}^\nu }{d \tau} \rangle = \langle \delta \tilde{\xi}^\mu  \partial_\alpha  \delta {\xi}^\nu \rangle \frac{d x^\alpha}{d \tau} =  \mathnormal{a}^{\mu \nu}_{\rho \alpha } \frac{d x^\alpha}{d \tau} \delta \overline{x}^\rho \tag{30}\\
& \mathnormal{h}^{\mu \nu}_{\rho} = \mathnormal{h}^{\nu \mu}_{\rho}, \mathnormal{k}^{\mu \nu}_{\rho \alpha \beta} = \mathnormal{k}^{\nu \mu}_{\rho \beta \alpha }  \tag{31}\\
& [{x}^\mu] = [\delta \overline{x}^\mu] = L, [\sigma] = L^{\frac{1}{2}}, [\delta \tilde{\xi}^\mu] = L^{\frac{1}{2}}, \\
& \ \ \ \ \ \ \ [\mathnormal{h}^{\mu \nu}_{\rho}] = L^0, [\mathnormal{k}^{\mu \nu}_{\rho \alpha \beta}] = L^{-2}, [\mathnormal{a}^{\mu \nu}_{\rho \alpha }] = L^{-1} \tag{32}
\end{align*}\\
We parametrize the two point functions, with the help of (15)-(17) above.  We continue treating space as isotropic, but we separate out the diffusive effects in space vs. time, and we do this by folding the square of the magnitude of diffusion into the definitions of the tensors $h, k, a$.\footnote{Thus, the dimension of each of these now becomes $[\mathnormal{h}^{\mu \nu}_{\rho}] = L^1, [\mathnormal{k}^{\mu \nu}_{\rho \alpha \beta}] = L^{-1}, [\mathnormal{a}^{\mu \nu}_{\rho \alpha }] = L^{0}$.}  We thus rewrite (15)-(17), to define the two point functions at some point in spacetime $x$ as follows:
\begin{align*}
&\sigma^2 \langle \delta \xi^\alpha_x \delta \xi^\beta_x \rangle|_{\alpha = \beta = i} \rightarrow \sigma^2 \eta^{i j}|_{i=j} \delta t \equiv h^{i j}_{0}|_{i=j} \delta t \tag{33}\\
&\sigma^2 \langle \delta \xi^\alpha_x \delta \xi^\beta_x \rangle|_{\alpha = \beta = 0} \rightarrow \sigma_\epsilon^2 \eta^{00} \sum_i \delta x^i \equiv h^{00}_{i} \delta x^i \tag{34}\\
&\sigma^2 \langle \partial_\mu \delta \xi^\alpha_x \partial_\nu \delta \xi^\beta_x \rangle|_{\alpha = \beta = i, \mu=\nu=0} \rightarrow \sigma^2 \eta^{i j}|_{i=j} \frac{2}{\delta t^2}\delta t \\
& \ \ \ \ \equiv k^{i j}_{000}|_{i=j} \delta t \tag{35}\\
& \sigma^2 \langle \partial_\mu \delta \xi^\alpha_x \partial_\nu \delta \xi^\beta_x \rangle|_{\alpha = \beta = i, \mu=\nu=k} \rightarrow \sigma^2 \eta^{i j}|_{i=j} \frac{2}{\delta x^k \delta t}\delta t \\
& \ \ \ \ \equiv k^{i j}_{0kl}|_{i=j,k=l} \delta t \tag{36} \\
&\sigma^2 \langle \partial_\mu \delta \xi^\alpha_x \partial_\nu \delta \xi^\beta_x \rangle|_{\alpha = \beta = 0, \mu=\nu=0} \rightarrow \sigma_\epsilon^2 \eta^{00} \frac{2}{\delta t \delta x^i}\delta x^i \\
& \ \ \ \ \equiv k^{00}_{i00} \delta x^i \tag{37}\\
& \sigma^2 \langle \partial_\mu \delta \xi^\alpha_x \partial_\nu \delta \xi^\beta_x \rangle|_{\alpha = \beta = 0, \mu=\nu=k} \rightarrow \sigma_\epsilon^2 \eta^{00} \frac{2}{\delta x^k \delta x^i}\delta x^i \\
& \ \ \ \ \equiv k^{00}_{ikl}|_{k=l} \delta x^i \tag{38} 
\end{align*} 
\text{}\\
From (33) and (34), we can directly identify tensor $h$, which is entirely defined by Markovian fields.  Tensor $k$ has both spacetime and Markovian field dependencies\footnote{Specifically, for tensor $k^{\alpha \beta}_{\rho\mu\nu}$, indecies $\alpha, \beta$, and $\rho$ reference Markovian fields or minimum length scale characteristic of the Markovian fields, while $\mu, \nu$ are spacetime indecies corresponding to the partial derivatives.} and can be generalized with an appropriate choice of diffusion parameters, shown below as $c$ and $c_\epsilon$.  Finally, to remove any confusion about index summation and to simplify notation, we will label the indecies explicitly in terms of $t,x,y,z$ from here on, to finally obtain:\footnote{Here we excluded the tensors which give zero contributions to the connection in both Minkowski and RW geometry.}
\begin{align*}
&h^{tt}_{x} = h^{tt}_{y} = h^{tt}_{z} = - \sigma_\epsilon^2 \tag{39} \\
& h^{xx}_{t} = h^{yy}_{t} = h^{zz}_{t} = \sigma^2 \tag{40} \\
& \mathnormal{k}^{t t}_{x t t}= \mathnormal{k}^{t t}_{x x x} = \mathnormal{k}^{t t}_{x y y} = \mathnormal{k}^{t t}_{x z z} = - 2 \mathnormal{c}_\epsilon \sigma_\epsilon^2 \tag{41}\\
& \mathnormal{k}^{x x}_{t t t}= \mathnormal{k}^{x x}_{t x x} = \mathnormal{k}^{x x}_{t y y} = \mathnormal{k}^{x x}_{t z z} = 2 \mathnormal{c} \sigma^2 \tag{42}
\end{align*}
Note that the diffusion parameters $c \sigma^2$ and $c_\epsilon \sigma_\epsilon^2$ effectively represent the diffusion acceleration rates in space and time, respectively.  Finally, it is not obvious in the Minkowski geometry, but ML under the RW metric shows us that all the tensors as well as the diffusion acceleration rates are defined not only by the nature of diffusion, but also by the geometry of choice.\\  
\text{}\\
Continuing to develop the math of ML under Minkowski metric, after some algebra we obtain
\begin{align*}
& \kappa^{t}_{t t} = \kappa^{t}_{x x} = \kappa^{t}_{y y} = \kappa^{t}_{z z} = - 3 \mathnormal{c} \sigma^2 \tag{43}\\
& \kappa^{x}_{t t} = \kappa^{x}_{x x} = \kappa^{x}_{y y} = \kappa^{x}_{z z} = \mathnormal{c}_{\epsilon} \sigma_\epsilon^2\tag{44}
\end{align*}\\
Finally, with the help of Einstein's Equations, our results correspond to a maximally symmetric universe:
\begin{align*}
& R^\alpha_{\sigma \mu \nu} = \kappa^\alpha_{\mu \beta} \kappa^\beta_{\nu \sigma} - \kappa^\alpha_{\nu \beta}\kappa^\beta_{\mu \sigma}, \\
& \ \ \ \ \ \ \ \ R_{\sigma \nu} = R^\rho_{\sigma \rho \nu} = \kappa^\alpha_{\alpha \beta} \kappa^\beta_{\nu \sigma} - \kappa^\alpha_{\nu \beta} \kappa^\beta_{\alpha \sigma} \tag{45} \\
& R_{t x} = R_{t y} = R_{t z}  = 3 c c_\epsilon \sigma \sigma_\epsilon, \ \ \ \ R_{t t} = 3 c_\epsilon^2 \sigma^4_\epsilon \tag{46}\\
& R_{x x} = R_{y y} = R_{z z}  =  9 c^2\sigma^4 +  2 c_\epsilon^2 \sigma^4_\epsilon  \approx 9 c^2 \sigma^4 \tag{47}\\
& R_{x y} = R_{y z} = R_{x z}  =  - \kappa^x_{y y} \kappa^y_{x x} = - c_\epsilon^2 \sigma^4_\epsilon \tag{48}
\end{align*}
\text{}\\
\text{}\\
\textbf{\emph{D. ML under RW geometry} }\\
\text{}\\
Under RW geometry we must adjust the explicit parametrization of the tensorial two-point functions to take into consideration the scale factor.  To do this we need to formulate how the tensors and diffusion parameters transform from Minkowski geometry to the RW geometry.  In the absence of Markovian fields this transformation is already well defined.  The metric transformation - from Minkowski geometry (denoted by $\alpha, \beta, \mu, \nu$, etc.) to the RW geometry (denoted by $\hat{\alpha}, \hat{\beta},\hat{ \mu}, \hat{\nu}$) - must satisfy
\begin{align*}
&g_{\hat{\alpha}\hat{\beta}} = \frac{\partial x^\alpha}{\partial x^{\hat{\alpha}}}\frac{\partial x^\beta}{\partial x^{\hat{\beta}}} \eta_{\alpha \beta} = (-1, a^2, a^2, a^2)\\
& \ \ \ \  \ \ \ \Rightarrow \ \ \ x^{\hat{i}} = \frac{1}{a} x^i\tag{49}
\end{align*}
The scale factor is only a function of time, giving us the simple relationship between the new (RW) and the old (Minkowski) spatial coordinates.\\
\text{}\\
\text{}\\
The tensor transformation in the absence of Markovian fields is also readily available,
\begin{align*}
&T^{\alpha \beta ...}_{\mu \nu ...} \longrightarrow T^{\hat{\alpha} \hat{\beta} ...}_{\hat{\mu} \hat{\nu} ...} =\frac{\partial x^{\hat{\alpha}} }{\partial x^{\alpha}}  \frac{\partial x^{\hat{\beta}}}{\partial x^{\beta}}...\frac{\partial x^\mu}{\partial x^{\hat{\mu}}} \frac{\partial x^\nu}{\partial x^{\hat{\nu}}}... T^{\alpha \beta ...}_{\mu \nu ...}\tag{50}
\end{align*}
But for a tensor which is exclusively a function of Markovian fields, we have:
\begin{align*}
&Q^{\alpha \beta ...}_{\mu \nu ...} \longrightarrow Q^{\hat{\alpha} \hat{\beta} ...}_{\hat{\mu} \hat{\nu} ...} =\\
& \ \ \ \ \frac{\partial \delta \xi^{\hat{\alpha}} }{\partial \delta \xi^{\alpha}}  \frac{\partial \delta \xi^{\hat{\beta}}}{\partial \delta \xi^{\beta}}...\frac{\partial \delta \xi^\mu}{\partial \delta \xi^{\hat{\mu}}} \frac{\partial \delta \xi^\nu}{\partial \delta \xi^{\hat{\nu}}}... Q^{\alpha \beta ...}_{\mu \nu ...}\tag{51}
\end{align*}
The last remaining task - for now - is to determine how the Markovian fields in the new metric relate to the Markovian fields in the old metric.  \\
\text{}\\
Let's consider the two-point function for Markovian fields in space, $\langle \delta \xi^x \delta \xi^x \rangle \sim h^{xx}_t$.  We already know that in the Minkowski geometry this two-point function scales with $\eta^{xx}$, which is just one.  But in the RW geometry each of the fields first obtains a factor of $\frac{1}{a}$ for 'residing' in space, and further the RW two point function obtains an additional factor of $g^{\hat{x}\hat{x}} = \frac{1}{a^2}$.\footnote{Note that this tells us that tensors defined entirely by Markovian fields experience an effective local geometry given by $g^{eff}_{\alpha \beta} = (-1, a^4, a^4, a^4, a^4)$.} Hence, we venture a guess:
\begin{align*}
&h^{\hat{x}\hat{x}}_{\hat{t}} = \frac{1}{a^4} h^{xx}_t \ \ \ \Rightarrow \ \ \ \frac{\partial \delta \xi^{\hat{x}} }{\partial \delta \xi^{x}} = \frac{1}{a^2} \tag{52}
\end{align*}
More generally, $h^{\alpha\beta}_\rho$ has indexes $\alpha, \beta$ which observe Markovian field transformations, and $\rho$ which is specific to how $\delta x^\rho$ transforms. $k^{\alpha\beta}_{\rho\mu\nu}$ has the same indexes as $h$, with the addition of $\mu, \nu$ which follow the ordinary rules of transformation.  By capturing the transformation of infinitesimal change in spacetime\footnote{Hence, minimum length.} from one geometry to another with some power of the scale factor, $a^z$ - where $z$ is to be defined,\footnote{Originally I have treated this transformation as equivalent to that of a Markovian field.  However, this was unnecessary in the absence of a gravitational source.} we are now ready to fully formulate the transformation of tensors $h,k$:
\begin{align*}
&h^{\alpha \beta}_{\rho } \longrightarrow h^{\hat{\alpha} \hat{\beta}}_{\hat{\rho}} =\frac{\partial \delta \xi^{\hat{\alpha}} }{\partial \delta \xi^{\alpha}}  \frac{\partial \delta \xi^{\hat{\beta}}}{\partial \delta \xi^{\beta}} \frac{\partial \delta x^\rho}{\partial \delta x^{\hat{\rho}}} h^{\alpha \beta}_{\rho}\tag{53}\\
&k^{\alpha \beta }_{\rho \mu \nu } \longrightarrow k^{\hat{\alpha} \hat{\beta}}_{\hat{\rho}\hat{\mu} \hat{\nu}} =\frac{\partial \delta \xi^{\hat{\alpha}} }{\partial \delta \xi^{\alpha}}  \frac{\partial \delta \xi^{\hat{\beta}}}{\partial \delta \xi^{\beta}} \frac{\partial \delta x^\rho}{\partial \delta x^{\hat{\rho}}} \frac{\partial x^\mu}{\partial x^{\hat{\mu}}} \frac{\partial x^\nu}{\partial x^{\hat{\nu}}} k^{\alpha \beta}_{\rho \mu \nu}\tag{54}\\
& \frac{\partial x^i}{\partial x^{\hat{i}}} = a, \ \ \ \frac{\partial \delta x^i}{\partial \delta x^{\hat{i}}} = a^z, \ \ \ \frac{\partial \delta \xi^{\hat{i}} }{\partial \delta \xi^{i}} = \frac{1}{a^2} \tag{55}
\end{align*}
From this point we work entirely in the RW geometry and can drop the use of 'hats' in the indecies.  The full set of tensors necessary for the formulation of Equations of Motion, Einstein, and Friedmann Equations is given by:
\begin{align*}
&h^{tt}_{x} = h^{tt}_{y} = h^{tt}_{z} = - \sigma^2_\epsilon a^z \tag{56}\\
& h^{xx}_{t} = h^{yy}_{t} = h^{zz}_{t} = \frac{\sigma^2}{a^4} \tag{57}\\
& \mathnormal{k}^{t t}_{x t t}= \mathnormal{k}^{t t}_{y t t}= \mathnormal{k}^{t t}_{z t t}= - 2 \mathnormal{c}_\epsilon \sigma_\epsilon^2 a^z \tag{58}\\
& \mathnormal{k}^{t t}_{x x x} = \mathnormal{k}^{t t}_{x y y} = \mathnormal{k}^{t t}_{x z z} = - 2 \mathnormal{c}_\epsilon \sigma_\epsilon^2 a^{2+z} \tag{59}\\
& \mathnormal{k}^{x x}_{t t t}= \mathnormal{k}^{yy}_{t t t}= \mathnormal{k}^{zz}_{t t t}= \frac{2 c \sigma^2}{a^4} \tag{60}\\
& \mathnormal{k}^{x x}_{t x x} = \mathnormal{k}^{x x}_{t y y} = \mathnormal{k}^{x x}_{t z z} = \frac{2 \mathnormal{c} \sigma^2 }{a^2}\tag{61}
\end{align*}
\text{}\\
During the formulation of the Equations of Motion in the absence of a gravitational source, all $a^z$ terms cancel.  Hence, RWML theory in the absence of a gravitational source is indifferent to the manner in which $\delta x^\rho$ transforms from Minkowski space to RW space.  To make all transformations consistent, we set $z=2$, to obtain:
\begin{align*}
& \kappa^{t}_{t t} = - 3 \frac{c \sigma^2}{a^2} \tag{62}\\
&\kappa^{t}_{x x} = \kappa^{t}_{y y} = \kappa^{t}_{z z} = - 3 \mathnormal{c} \sigma^2 \tag{63}\\
& \kappa^{x}_{t t} = \kappa^{y}_{t t} =\kappa^{z}_{t t} =c_\epsilon \sigma^2_\epsilon \tag{64}\\
&\kappa^{x}_{x x} = \kappa^{x}_{y y} = \kappa^{x}_{z z} = \mathnormal{c}_\epsilon \sigma^2_\epsilon a^2 \tag{65}\\
& \Sigma^t_{xx}=  \Sigma^t_{yy}=  \Sigma^t_{zz}=  - \frac{3}{2}\frac{\kappa \sigma^2}{a^2} \tag{66} \\
& \Sigma^x_{xx}=  \Sigma^x_{yy}=  \Sigma^x_{zz}= - \frac{\sigma^2_\epsilon}{2} (a \ddot{a} + \dot{a}^2) \tag{67}\\
& \Gamma^t_{xx} = \Gamma^t_{yy} = \Gamma^t_{zz} = a \dot{a} \tag{68}\\
& \Gamma^x_{tx} = \Gamma^y_{ty} = \Gamma^z_{tz} = \frac{\dot{a}}{a} \tag{69}
\end{align*}
The elements of the new connection are now given by:
\begin{align*}
&\tilde{\Gamma}^t_{tt} = 3 \frac{c \sigma^2}{a^2} \tag{70}\\
&\tilde{\Gamma}^t_{xx} = \tilde{\Gamma}^t_{yy} = \tilde{\Gamma}^t_{zz} = a \dot{a} + 3 c \sigma^2 + \frac{3}{2}\frac{\kappa \sigma^2}{a^2} \tag{71} \\
& \tilde{\Gamma}^x_{tt} = \tilde{\Gamma}^y_{tt} =\tilde{\Gamma}^z_{tt} = - c_\epsilon \sigma^2_\epsilon \tag{72}\\
&\tilde{\Gamma}^x_{tx} = \tilde{\Gamma}^y_{ty} = \tilde{\Gamma}^z_{tz} = \frac{\dot{a}}{a} \tag{73}\\
&\tilde{\Gamma}^x_{xx} = \tilde{\Gamma}^x_{yy} =\tilde{\Gamma}^x_{zz} = - c_\epsilon \sigma^2_\epsilon a^2 + \frac{\sigma^2_\epsilon}{2} (a \ddot{a} + \dot{a}^2) \tag{74}
\end{align*}
\text{}\\
As already discussed in VII.B., we must consider how the spatial diffusive parameter $\sigma$ transforms from Minkowski to RW geometry.  It is useful to note that $\sigma$ has the same dimension as $\xi^\alpha$: $[\sigma]=L^{\frac{1}{2}}$, and thus we would expect that it transforms in the same manner.  We can define its parameter transformation as $\sigma =\sigma_{Minkowski} \ \ \ \longrightarrow \ \ \ \sigma =\frac{\sigma_{Minkowski}}{a^y}$, where we still need to obtain the power of the scale factor, $y$.  If we are correct about our intuitive guess with regards to ML field transformations, we ought to find that $y=2$.\\
\text{}\\
The Friedmann Equations show us that a physical universe with a positive Hubble parameter for a slow moving $D$ relative to $H$ ($|\dot{D}|<< |\dot{H}|$), is not possible for $y<2$.  We thus set $y=2$, the lowest possible value in obtaining a physical universe which matches experimental observations, giving us $\sigma \sim \frac{1}{a^2}$.\\
\text{}\\
Equipped with the above, we can formulate the Reimann tensor to ultimately obtain the components of the Ricci tensor and the Ricci scalar:
\begin{align*}
& R_{tt} = -3 \frac{\ddot{a}}{a} + 9 \frac{\dot{a}}{a} \frac{c \sigma^2}{a} + O(c_\epsilon \sigma_\epsilon^4) \tag{75}\\
& R_{xx} = R_{yy} = R_{zz} = \\
& \ \ \ \  a \ddot{a} + 2 \dot{a}^2 + 2 \kappa - 6 \frac{\dot{a}}{a} c \sigma^2  + 9 \frac{c^2 \sigma^4}{a^2}\\
& \ \ \ \  + O(\frac{\kappa}{a^2}) + O(c_\epsilon \sigma_\epsilon^4) \tag{76}\\
& R = 6 (\frac{\ddot{a}}{a} + (\frac{\dot{a}}{a})^2) - 27 \frac{\dot{a}}{a} \frac{c \sigma^2}{a^2} + 27 \frac{c^2 \sigma^4}{a^4} \\
& \ \ \ \ + O(\frac{\kappa}{a^2}) + O(c_\epsilon \sigma_\epsilon^4)\tag{77}
\end{align*}
\text{}\\
\textbf{\emph{E. Expansion acceleration}}\\
\text{}\\
The new RWML connection allows us to obtain equations of motion for each species.  In fact, there are two such equations, one giving us the evolution of species-specific energy density through time, and the other giving us the gradient of pressure.  In an isotropic vacuum we expect the gradient of pressure to be zero. This second equation for both RW theory without diffusion and ML theory in Minkowski geometry is trivial.  But here we learn that, in fact, temporal diffusion cannot be assumed away.  While it appears to scale as extremely small, it is the driving force behind the positive acceleration of the universe expansion.\\
\text{}\\
Assuming an isotropic universe, the pressure-specific equation of motion for RWML gives us:
\begin{align*}
& c_\epsilon a^2 = \frac{2 w}{1+4w} (a \ddot{a} + \dot{a}^2) \tag{78} 
\end{align*}
With this we obtain Equation (4) for the case of a vacuum-dominated universe ($w=-1$).  Note that matter ($w=0$) does not participate in temporal acceleration.\\
\text{}\\
\text{}\\
\text{}\\
\text{}\\
\textbf{\emph{F. Species-specific energy density}}\\
\text{}\\
The more complete equation\footnote{See the Appendix for even more detail.} describing the species-specific energy density evolution is given by $\nabla_\mu T^{\mu t}$:
\begin{align*}
& \frac{\dot{\rho}_w}{{\rho}_w} = - 3 \{(1+w) \frac{\dot{a}}{a} + \frac{c_w \sigma_w^2}{a^2} (2 + 3w(1 + \frac{\kappa}{2 c a^2}) \} \\
& \ \ \ \ + O((\frac{\dot{a}}{a})^2 \frac{c_w \sigma_w^2}{a^2})\tag{79} 
\end{align*}
\text{}\\ 
\text{}\\
\textbf{\emph{G. RWML Friedmann Equations}}\\
\text{}\\
The first and second Friedmann Equations in RWML theory are given by:\\
\begin{align*}
& \frac{8 \pi G}{3}\rho = (\frac{\dot{a}}{a})^2 - \frac{3}{2} \frac{\dot{a}}{a} \frac{c \sigma^2}{a^2} + \frac{9}{2} \frac{c^2 \sigma^4}{a^4} + O(\frac{\kappa}{a^2}) \\
& \ \ \ \ \ \ \ \ \ \ \ \ \ \approx H^2 - \frac{3}{2} H D + \frac{9}{2}  D^2 \tag{80} \\
& - {8 \pi G}p = 2 \frac{\ddot{a}}{a} + (\frac{\dot{a}}{a})^2  - \frac{15}{2} \frac{\dot{a}}{a} \frac{c \sigma^2}{a^2} + \frac{9}{2} \frac{c^2 \sigma^4}{a^4} \\
& \ \ \ \ \ \ \ \ \ \ \ \ \ \ \ \ \ \ \ \ \ \ \ \ \ \ + O(\frac{\kappa}{a^2}) \\
& \ \ \ \ \ \ \ \ \ \ \ \ \ \approx (2f + 1) H^2 - \frac{15}{2} H D + \frac{9}{2} D^2  \tag{81}
\end{align*}
\text{}\\
With a particular choice for $H$ and $f$, we can solve for $w$ ($p=w \rho$).  For example, setting $H = \frac{\dot{a}}{a} = \frac{3}{4} \frac{c \sigma^2}{a^2}$, the value which minimizes the total energy density for a slow moving $D$, and choosing $f = \frac{1}{2}$, we obtain $w=0$.  Under RWML theory, Friedmann Equations for a fast moving Hubble parameter relative to the diffusive parameter describe a matter-dominated universe. \\
\text{}\\
Alternatively we can consider the case where the Hubble parameter does not change in value. This occurs when $\dot{H} = (f-1) H^2 = 0$, thus when the acceleration parameter equals one. In this case the energy density is minimized when $D = \frac{H}{6}$, giving us $w = -\frac{5}{7} \approx -0.7$, the universe with the effective equation of state as we understand it to be presently, mostly a $2:1$ mix of vacuum and matter.  \\
\text{}\\
In the last example here, we can consider a static universe described by the Friedmann Equations of RWML theory, with both zero velocity and acceleration.  Just as in the case of RW theory, the only species which can accommodate this scenario is curvature, with $w=-\frac{1}{3}$.\\
\text{}\\ 
\text{}\\
\textbf{\emph{H. Effective universe equation of state and deceleration parameter}}\\
\text{}\\
If we assume that $D$ is approximately constant, we can minimize the energy density relative to $H$, to obtain a minimum energy density defined by $D^2$:
\begin{align*}
&\frac{\partial \rho}{\partial H} = 0 \ \ \ \Rightarrow \ \ \ H = \frac{3}{4} D, \ \ \ \frac{8\pi G}{3}\rho = \frac{63}{16} D^2,\\
& \ \ \ \  \ \ \ f = \frac{1}{2}(1 - 21 w) \tag{82} \\
&f = 0.5 \ \ \ \Rightarrow \ \ \ w = 0 \tag{83} 
\end{align*}
\text{}\\
On the other hand, for $H$ approximately constant, we can minimize the energy density relative to $D$:\\
\begin{align*}
&\frac{\partial \rho}{\partial D} = 0 \ \ \ \Rightarrow \ \ \ D = \frac{1}{6} H, \ \ \ \frac{8\pi G}{3}\rho = \frac{7}{8} H^2, \\
& \ \ \ \ \ \ \ f = \frac{1}{16}(1 - 21 w) \tag{84} \\
& f = 1 \ \ \ \Rightarrow \ \ \ w \approx -\frac{5}{7} \approx -0.7\tag{85} 
\end{align*}
However, we can be more precise.  Both $H$ and $D$ are functions of time as follows:
\begin{align*}
& \frac{\ddot{a}}{a} \equiv f (\frac{\dot{a}}{a})^2 \ \ \ \Rightarrow \ \ \ \dot{H} = -(1-f) H^2 \tag{86} \\
&\sigma \sim \frac{1}{a^2} \ \ \ \Rightarrow \ \ \ D \sim \frac{1}{a^6} \ \ \ \Rightarrow \ \ \ \dot{D} =  - 6 H D \tag{87} \\
&\frac{8\pi G}{3} \dot{\rho} = - 36H (D^2 - \frac{1}{36} (7 - f) HD \\
& \ \ \ \ \ \ \ + \frac{1}{27} (1 - f) H^2) \tag{88} \\
& \dot{\rho} = 0 \Rightarrow  D^{\pm} \equiv \frac{H}{72}  \{ 7 - f  \pm \sqrt{f^2 + 178 f - 143} \} \tag{89} \\
& f \geqslant 0.7998 \tag{90}\\
& f \approx 1 \ \ \ \Rightarrow \ \ \ D^+ \approx \frac{H}{6}, \ \ \ w \approx -0.7 \tag{91} \\
& f \rightarrow  \infty \Rightarrow D^+ \rightarrow \frac{4}{3} H,  \ D^- \rightarrow  (-\infty) H  \tag{92} 
\end{align*}
We see that for the acceleration parameter in the neighborhood of $1$ we obtain a universe solidly in the mostly vacuum-dominated region and well approximated by the RW theory.  We also see that for the energy density to have a well defined minimum with a real density parameter, the acceleration parameter must be greater than or equal to $0.7998$.\\
\text{}\\
Finally, consider how the choice of the local minimum determines whether the universe is flat, open, or closed, as defined by the relative value of energy density to the critical energy density:
\begin{align*}
&\rho_c \equiv \frac{3 H^2}{8 \pi G}, \ \ \ \rho_d \equiv \frac{3 D^2}{8 \pi G}, \ \ \ \Omega_d \equiv \frac{\rho_d}{\rho_c} = \frac{D^2}{H^2}  \tag{93}\\
&\rho = \rho_c - \frac{3}{2} \sqrt{\rho_d \rho_c} + \frac{9}{2} \rho_d + \frac{3 \kappa}{8 \pi G a^2}\tag{94} \\
&\Omega -1 = \frac{9}{2} \Omega_d - \frac{3}{2} \sqrt{\Omega_d} + \frac{\kappa}{H^2 a^2} \tag{95} 
\end{align*}
Note that we can now have a geometrically flat universe with $\kappa \approx 0$ along with a closed or open universe as a function of the diffusive energy density relative to the critical density.  For a geometrically flat universe we have:
\begin{align*}
&\Omega > 1 \ \ \ \Rightarrow \ \ \ D > \frac{1}{3} H, \ \ \ f > 3  \tag{96} \\
&\Omega < 1 \ \ \ \Rightarrow \ \ \ D < \frac{1}{3} H, \ \ \ f < 3   \tag{97} 
\end{align*}
\text{}\\
\textbf{{VIII. ACKNOWLEDGEMENTS}}\\
\text{}\\
I would like to thank Maja Buri\'{c}, Adam Peterson, and Pavel Wiegmann for valuable discussions and extremely thoughtful questions and comments.  I would also like to thank Rick Cavanaugh and the CMS Collaboration for their encouragement and support.  While this paper has greatly benefited from such assistance, any remaining errors are my own.\\
\text{}\\
\text{}\\
\textbf{{APPENDIX A: ML CONNECTION}}\\
\begin{align*} 
& I = - \frac{1}{2} \int d\tau \frac{d x^\mu}{d \tau} g_{\mu \nu} \frac{d x^\nu}{d \tau}, \\
& \ \ \ \ \ x^\mu \rightarrow x^\mu + \delta \overline{x}^\mu + \sigma \delta {\xi}^\mu, \delta \xi^\mu \sim \mathcal{N}(0,\sqrt{\delta t}) \tag{98}\\
& I \rightarrow - \frac{1}{2} \int d\tau \{ \frac{d(x^\mu + \delta \overline{x}^\mu + \sigma \delta \tilde{\xi}^\mu)}{d \tau} (g_{\mu \nu} + (\partial_\rho g_{\mu \nu})\\
& \ \ \ \ \ \ \ \ \ \ \ \ \ \ \ (\delta \overline{x}^\rho + \sigma \delta \tilde{\xi}^\rho) \\
& \ \ \ \ \ + \frac{1}{2} (\partial_\alpha \partial_\beta g_{\mu \nu}) \sigma^2 \mathnormal{h}^{\alpha \beta}_\rho \delta \overline{x}^\rho )\frac{d(x^\nu + \delta \overline{x}^\nu + \sigma \delta \tilde{\xi}^\nu)}{d \tau} \} \tag{99}
\end{align*}
\begin{align*}
&\langle \delta I \rangle = - \frac{1}{2} \int d\tau \{ (\partial_\alpha g_{\mu\nu}) \delta x^\alpha \frac{dx^\mu}{d\tau}\frac{dx^\nu}{d\tau}\\
& \ \ \ \ \ \  + \frac{1}{2} \sigma^2 (\partial_\alpha\partial_\beta g_{\mu\nu}) \langle \delta \xi^\alpha \delta \xi^\beta \rangle \frac{dx^\mu}{d\tau}\frac{dx^\nu}{d\tau} \\
&  \ \ \ -\partial x^\nu (\partial_\alpha g_{\mu\nu}) \frac{dx^\alpha}{d\tau}\frac{dx^\mu}{d\tau} - \partial x^\nu g_{\mu\nu}\frac{d^2 x^\mu}{d \tau^2} \\
& \ \ \ \ \ \ -\partial x^\mu (\partial_\alpha g_{\mu\nu}) \frac{dx^\alpha}{d\tau}\frac{dx^\nu}{d\tau} - \partial x^\mu g_{\mu\nu}\frac{d^2 x^\nu}{d \tau^2} \\
& \ \ \ + \sigma \langle \delta \xi^\beta (\partial_\alpha \sigma \xi^\nu) \rangle (\partial_\beta g_{\mu\nu}) \frac{dx^\alpha}{d\tau}\frac{dx^\mu}{d\tau} \\
& \ \ \ \ \ \ + \sigma \langle \delta \xi^\beta (\partial_\alpha \sigma \xi^\mu) \rangle (\partial_\beta g_{\mu\nu}) \frac{dx^\alpha}{d\tau}\frac{dx^\nu}{d\tau}\\
& \ \ \ + \langle (\partial_\alpha \sigma \xi^\mu)(\partial_\beta \sigma \xi^\nu) \rangle g_{\mu\nu}\frac{dx^\alpha}{d\tau}\frac{dx^\beta}{d\tau} \} \tag{100}
\end{align*}
\begin{align*}
&\langle \delta I \rangle = \int d\tau g_{\mu\rho} \delta x^\rho \{ \frac{d^2 x^\mu}{d \tau^2} \\
& \ \ \ \ \ \ + \frac{dx^\alpha}{d\tau}\frac{dx^\beta}{d\tau} [\frac{1}{2} g^{\mu \sigma} (\partial_\alpha g_{\beta \sigma} + \partial_\beta g_{\sigma \alpha} - \partial_\sigma g_{\alpha \beta }) \\
& \ \ \ \ \ \ - \frac{1}{2} g^{\mu \delta} (\frac{1}{2} \sigma^2  \mathnormal{h}^{\eta \sigma}_{\delta} (\partial_\eta \partial_\sigma g_{\alpha \beta}) +  (\partial_\alpha \sigma^2) (\partial_\sigma g_{\eta \beta}) \mathnormal{h}^{\eta \sigma}_{\delta} \\
& \ \ \ \ \ \ + (\partial_\alpha \sigma) (\partial_\beta \sigma) \mathnormal{h}^{\eta \sigma}_{\delta} g_{\eta \sigma} +  2 \sigma^2 a^{\sigma \eta}_{\delta \alpha} (\partial_\sigma g_{\eta \beta}) \\
& \ \ \ \ \ \ +  (\partial_\alpha \sigma^2) a^{\eta \sigma }_{\delta \beta} g_{\eta \sigma } +   \sigma^2 \mathnormal{k}^{\eta \sigma}_{\delta \alpha \beta} g_{\eta \sigma }) ]\}  \tag{101}
\end{align*}
\text{}\\
\textbf{{APPENDIX B: TWO-POINT FUNCTIONS IN RW GEOMETRY}}\\
\begin{align*}
& \frac{\partial x^i}{\partial x^{\hat{i}}} = a, \ \ \ \frac{\partial \delta x^i}{\partial \delta x^{\hat{i}}} = a^z, \ \ \ \frac{\partial \delta \xi^{\hat{i}} }{\partial \delta \xi^{i}} = \frac{1}{a^2} \tag{102}\\
&h^{\hat{\alpha} \hat{\beta}}_{\hat{\rho}} =\frac{\partial \delta \xi^{\hat{\alpha}} }{\partial \delta \xi^{\alpha}}  \frac{\partial \delta \xi^{\hat{\beta}}}{\partial \delta \xi^{\beta}} \frac{\partial \delta x^\rho}{\partial \delta x^{\hat{\rho}}} h^{\alpha \beta}_{\rho}\tag{103}\\
&\Rightarrow \ \ \ h^{\hat{x} \hat{x}}_{\hat{t}} = \frac{1}{a^2} \frac{1}{a^2} \sigma^2 = \frac{\sigma^2}{a^4}, \\
& \ \ \ \ \ \ \ \ h^{\hat{t} \hat{t}}_{\hat{x}} =  a^z (-\sigma_\epsilon^2) = - \sigma_\epsilon^2 a^z\tag{104}\\
&k^{\hat{\alpha} \hat{\beta}}_{\hat{\rho}\hat{\mu} \hat{\nu}} =\frac{\partial \delta \xi^{\hat{\alpha}} }{\partial \delta \xi^{\alpha}}  \frac{\partial \delta \xi^{\hat{\beta}}}{\partial \delta \xi^{\beta}} \frac{\partial \delta x^\rho}{\partial \delta x^{\hat{\rho}}} \frac{\partial x^\mu}{\partial x^{\hat{\mu}}} \frac{\partial x^\nu}{\partial x^{\hat{\nu}}} k^{\alpha \beta}_{\rho \mu \nu}\tag{105}\\
&\Rightarrow \ \ \ k^{\hat{x} \hat{x}}_{\hat{t}\hat{t} \hat{t}} = \frac{1}{a^2} \frac{1}{a^2} 2 c \sigma^2 = \frac{2 c \sigma^2}{a^4},  \\
& \ \ \ \ \ \ \ \ k^{\hat{x} \hat{x}}_{\hat{t}\hat{y} \hat{y}} = \frac{1}{a^2} \frac{1}{a^2} a a 2 c \sigma^2 = \frac{2 c \sigma^2}{a^2} \tag{106}\\
&\Rightarrow \ \ \ k^{\hat{t} \hat{t}}_{\hat{x}\hat{t} \hat{t}} = a^z (- 2 c_\epsilon \sigma_\epsilon^2) = - 2 c_\epsilon \sigma_\epsilon^2 a^z,  \\
& \ \ \ \ \ \ \ \ k^{\hat{t} \hat{t}}_{\hat{x}\hat{y} \hat{y}} = a^z a a (- 2 c_\epsilon \sigma_\epsilon^2) = - 2 c_\epsilon \sigma_\epsilon^2 a^{2+z} \tag{107}
\end{align*}
\text{}\\
\textbf{{APPENDIX C: CONNECTION TENSORS IN RW GEOMETRY}}\\
\text{}\\
To make the derivations clearer, we work with the two-point functions $h$ and $k$  prior to folding $\sigma$ or $\sigma_\epsilon$ into their definitions:
\begin{align*}
& \partial_x^2 g |_{r=0} = 2 \kappa (0, a^2), \ \ \ \partial_t^2 g |_{r=0} = 2 (0, \dot{a}^2 + a \ddot{a}), \\
& \ \ \ \ \ \ \ \ \ \ \ \partial_x g |_{r=0} = 0  \tag{108}\\
& \sigma \sim \frac{1}{a^{y}}, \ \ \ \partial_t \sigma = -y \sigma \frac{\dot{a}}{a}, \ \ \ \partial_t \sigma^2 = -2y \sigma^2 \frac{\dot{a}}{a}, \\
& \ \ \ \ \ \ \ \ \ \ \ \partial_t \sigma_\epsilon =  0 \tag{109}
\end{align*}
\begin{align*}
& \Sigma ^{\mu}_{\alpha \beta} \equiv \frac{1}{4} g^{\mu \delta} \sigma^2  \mathnormal{h}^{\eta \sigma}_{\delta} (\partial_\eta \partial_\sigma g_{\alpha \beta})\\
& \ \ \ \ \ \ \ \  + \frac{1}{2} g^{\mu \delta} (\partial_\alpha \sigma^2) (\partial_\sigma g_{\eta \beta}) \mathnormal{h}^{\eta \sigma}_{\delta}  \\
& \ \ \ \ \ \ +\frac{1}{2} g^{\mu \delta} (\partial_\alpha \sigma) (\partial_\beta \sigma) \mathnormal{h}^{\eta \sigma}_{\delta} g_{\eta \sigma} \tag{110}\\
&\Rightarrow \Sigma^{t}_{\alpha \beta} = \frac{3}{4} g^{tt} \sigma^2 \mathnormal{h}^{xx}_{t} (\partial_x^2 g_{\alpha \beta}) \\
& \ \ \ \ \ \ \ \ + \frac{3}{2} g^{tt} (\partial_\alpha \sigma) (\partial_\beta \sigma) \mathnormal{h}^{xx}_{t} g_{xx} \tag{111}\\
& \Rightarrow \Sigma^{t}_{tt} = - \frac{3}{2} y^2 \frac{\sigma^2}{a^2} (\frac{\dot{a}}{a})^2 = O(\frac{\sigma^2}{a^2} H^2), \\
& \ \ \ \ \ \ \ \ \ \ \ \Sigma ^{t}_{xx} = - \frac{3}{2} \frac{\kappa \sigma^2}{a^2} = O(\sigma^2 \frac{\kappa}{a^2}) \tag{112}
\end{align*}
\begin{align*}
&\Rightarrow \Sigma^{x}_{\alpha \beta} = \frac{1}{4} g^{xx} \sigma_\epsilon^2 \mathnormal{h}^{tt}_{x} (\partial_t^2 g_{\alpha \beta}) + 0 + 0 \tag{113}\\
& \ \ \ \Rightarrow \Sigma ^{x}_{tt} = 0, \ \ \ \Sigma^{x}_{yy} = -\frac{1}{2} \sigma_\epsilon^2  a^z ( \frac{\ddot{a}}{a} + (\frac{\dot{a}}{a})^2) \tag{114}\\
&\kappa^{\mu}_{\alpha \beta} \equiv \frac{1}{2} g^{\mu \delta} \sigma^2 \mathnormal{k}^{\eta \sigma}_{\delta \alpha \beta} g_{\eta \sigma } \tag{115} \\
&\Rightarrow \kappa^{t}_{tt} = \frac{3}{2} g^{tt} \sigma^2 \mathnormal{k}^{xx}_{ttt} g_{xx} = -3\frac{c \sigma^2}{a^2}, \\
& \ \ \ \ \ \ \ \ \ \ \ \kappa^{t}_{xx} = \frac{3}{2} g^{tt} \sigma^2 \mathnormal{k}^{yy}_{txx} g_{yy} = -3{c \sigma^2}\ \ \  \tag{116} \\
&\Rightarrow \kappa^{x}_{tt} = \frac{1}{2} g^{xx} \sigma_\epsilon^2 \mathnormal{k}^{tt}_{xtt} g_{tt} = \frac{c_\epsilon \sigma_\epsilon^2 a^z}{a^2},\\
& \ \ \ \ \ \ \ \  \ \ \ \kappa^{x}_{yy} = \frac{1}{2} g^{xx} \sigma_\epsilon^2 \mathnormal{k}^{tt}_{xyy} g_{tt} = c_\epsilon \sigma_\epsilon^2 a^z\ \ \  \tag{117} \\
\end{align*}
\begin{align*}
& \mathcal{A}^{\mu}_{\alpha \beta} \equiv \frac{1}{2} g^{\mu \delta} \sigma^2 a^{\sigma \eta}_{\delta \alpha} (\partial_\sigma g_{\eta \beta}) \ \ \ \Rightarrow \ \ \ \mathcal{A}|_{r=0} = 0 \tag{118}\\
& \Gamma^{\mu}_{\alpha \beta} = \frac{1}{2} g^{\mu \sigma} (\partial_\alpha g_{\beta \sigma} + \partial_\beta g_{\sigma \alpha} - \partial_\sigma g_{\alpha \beta }) \\
& \ \ \ \ \ \ \ \ \ \ \ \Rightarrow \ \ \ \Gamma^{t}_{xx} = a \dot{a}, \ \ \ \Gamma^{x}_{xt} = \frac{\dot{a}}{a} \tag{119}\\
&\tilde{\Gamma}^{\mu}_{\alpha \beta} = {\Gamma}^{\mu}_{\alpha \beta} - {\kappa}^{\mu}_{\alpha \beta} - \mathcal{A}^{\mu}_{\alpha \beta} - {\Sigma}^{\mu}_{\alpha \beta} \tag{120}\\
&\Rightarrow \tilde{\Gamma}^{t}_{tt} = 3\frac{c \sigma^2}{a^2} (1 + \frac{y^2}{2 c}(\frac{\dot{a}}{a})^2)  = 3\frac{c \sigma^2}{a^2} + O(\frac{\sigma^2}{a^2} H^2) \\
& \ \ \ \ \ \ \ \ \approx 3\frac{c \sigma^2}{a^2}, \tag{121}\\
&\Rightarrow \tilde{\Gamma}^{t}_{xx} = a \dot{a} +  3\frac{c \sigma^2}{a^2} + \frac{3}{2} \frac{\kappa \sigma^2}{a^2} = a \dot{a} +  3\frac{c \sigma^2}{a^2} + O(\sigma^2 \frac{\kappa }{a^2}) \\
& \ \ \ \ \ \ \ \ \approx a \dot{a} +  3\frac{c \sigma^2}{a^2}\tag{122}\\
&\Rightarrow \tilde{\Gamma}^{x}_{tt} = - \frac{c_\epsilon \sigma_\epsilon^2 a^z}{a^2},  \ \ \ \tilde{\Gamma}^{x}_{xt} = \frac{\dot{a}}{a},\\
& \ \ \ \ \ \ \ \  \ \ \ \tilde{\Gamma}^{x}_{yy} = \frac{1}{2} \sigma_\epsilon^2  a^z ( \frac{\ddot{a}}{a} + (\frac{\dot{a}}{a})^2) - c_\epsilon \sigma_\epsilon^2 a^z \tag{123}
\end{align*}
\text{}\\
\textbf{{APPENDIX D: RWML EQUATIONS OF MOTION}}\\
\text{}\\
Taking $z=2$ from here on, we have
\begin{align*}
& \nabla_\mu{T^{\mu\nu}} = \partial_\mu T^{\mu \nu} + \tilde{\Gamma}^{\alpha}_{\alpha\beta} T^{\beta \nu} + \tilde{\Gamma}^{\nu}_{\alpha\beta} T^{\alpha \beta} = 0, \\
& \ \ \ \ \ \ \ \ T_{\mu\nu} = \rho(1, w a^2, w a^2, w a^2) \tag{124}\\
&\partial_t T^{tt} + (2 \tilde{\Gamma}^{t}_{tt} + 3 \tilde{\Gamma}^{x}_{xt}) T^{tt} + 3 \tilde{\Gamma}^{t}_{xx} T^{xx} = 0 \tag{125}\\
& \Rightarrow \partial_t\rho = - 3 \rho \{(1+w) \frac{\dot{a}}{a} + \\
& \ \ \ \ \ \frac{c \sigma^2}{a^2} (2 + \frac{2}{c} (\frac{\dot{a}}{a})^2 + 3w(1 + \frac{\kappa}{2 c a^2}) \}\tag{126} \\
&4 \tilde{\Gamma}^{x}_{xx} T^{xx} +  \tilde{\Gamma}^{x}_{tt} T^{tt} = 0 \tag{127}\\
& \Rightarrow \ \ \ c_\epsilon (1+4w) = 2 w (\frac{\ddot{a}}{a} + (\frac{\dot{a}}{a})^2)\tag{128} 
\end{align*}
\textbf{{APPENDIX E: RWML Ricci tensor, scalar, and Friedmann Equations}}
\begin{align*}
&R_{\mu\nu} = \partial_\alpha \tilde{\Gamma}^\alpha_{\mu\nu} - \partial_\nu \tilde{\Gamma}^\alpha_{\alpha\mu} + \tilde{\Gamma}^\alpha_{\alpha\beta} \tilde{\Gamma}^\beta_{\mu\nu}- \tilde{\Gamma}^\alpha_{\nu\beta} \tilde{\Gamma}^\beta_{\alpha\mu} \tag{129} 
\end{align*}
\begin{align*}
&\Rightarrow  \ R_{tt} = \partial_\alpha \tilde{\Gamma}^\alpha_{tt} - \partial_t \tilde{\Gamma}^\alpha_{\alpha t} + \tilde{\Gamma}^\alpha_{\alpha\beta} \tilde{\Gamma}^\beta_{tt}- \tilde{\Gamma}^\alpha_{t \beta} \tilde{\Gamma}^\beta_{\alpha t} \\
& \ \ \ \ \ \ \ \ \ \ = R_{tt}^{RW} + 3 \tilde{\Gamma}^x_{xt} \tilde{\Gamma}^t_{tt} + 3 \tilde{\Gamma}^x_{xx} \tilde{\Gamma}^x_{tt} \\
&  \ \ \ \ \ \ \ \ \ \ = -3 \frac{\ddot{a}}{a} + 9 \frac{\dot{a}}{a} \frac{c \sigma^2}{a^2} + O(\sigma_\epsilon^4)\tag{130} \\
&\Rightarrow \ R_{xx} = \partial_\alpha \tilde{\Gamma}^\alpha_{xx} - \partial_x \tilde{\Gamma}^\alpha_{\alpha x} + \tilde{\Gamma}^\alpha_{\alpha\beta} \tilde{\Gamma}^\beta_{xx}- \tilde{\Gamma}^\alpha_{x \beta} \tilde{\Gamma}^\beta_{\alpha x} \\
&  \ \ \ \ \ \ \ \ \ \ = R_{xx}^{RW} - (\partial_t + \tilde{\Gamma}^y_{yt}) (\kappa^t_{xx} + \Sigma^t_{xx}) \\
& \ \ \ \ \ + \tilde{\Gamma}^t_{tt} \tilde{\Gamma}^t_{xx} + 2 \tilde{\Gamma}^x_{xx} \tilde{\Gamma}^x_{xx} \\
&  \ \ \ \ \ \ \ \ \ \ = a \ddot{a} + 2 (\dot{a})^2 + 2 \kappa - 6 (y-1) \frac{\dot{a}}{a} c \sigma^2 \\
& \ \ \ \ \ - \frac{3}{2} (2y+1) \frac{\dot{a}}{a}  \frac{\kappa \sigma^2}{a^2} + \frac{9}{2} \frac{c^2 \sigma^4}{a^2} (2 + \frac{\kappa }{c a^2}) \\
&  \ \ \ \ \ \ \ \ \ \ \ \ \ \ \ \ \ \ + O(c_\epsilon \sigma_\epsilon^2)\tag{131} 
\end{align*}
Setting $y=2$, the minimum value to give us a universe with an energy density which has a local minimum at a positive Hubble parameter, we obtain the Ricci scalar, and Friedmann Equations:
\begin{align*}
&R = g^{\mu\nu} R_{\mu\nu} = -R_{tt} + \frac{3}{a^2} R_{xx} = \\
& \ \ \ \ \ 6 (\frac{\ddot{a}}{a} + (\frac{\dot{a}}{a})^2 + \frac{\kappa}{a^2}) - 27 \frac{\dot{a}}{a} \frac{c \sigma^2}{a^2} + 27 \frac{c^2 \sigma^4}{a^4} \tag{132} \\
&R_{\mu\nu} - \frac{1}{2}R g_{\mu\nu} = 8\pi G T_{\mu\nu} \tag{133} \\
&\Rightarrow \ R_{tt} + \frac{1}{2}R = 8\pi G \rho, \ \ \ R_{xx} - \frac{1}{2}R a^2 = 8\pi G \rho w a^2 \tag{134} \\
&\frac{\ddot{a}}{a} \approx f (\frac{\dot{a}}{a})^2, \ \ \ H \equiv \frac{\dot{a}}{a}, \ \ \ D \equiv \frac{c \sigma^2}{a^2}  \tag{135} \\
&\Rightarrow \frac{8\pi G}{3} \rho = H^2  - \frac{3}{2} H D + \frac{9}{2} D^2 + O(\frac{\kappa}{a^2}) + O(\sigma_\epsilon^4)\tag{136} \\
&\Rightarrow - \frac{8\pi G}{3} w \rho = \frac{1}{3} (2 f + 1) H^2 - \frac{5}{2} H D + \frac{3}{2} D^2 \\
& \ \ \ \ \ + O(\frac{\kappa}{a^2}) + O(\sigma_\epsilon^4) \tag{137} 
\end{align*}
where $f = -q$ is the acceleration parameter, the negative of the typically used deceleration parameter $q$.\\
\text{}\\
\nocite{*}
%\printbibliography
\textbf{{BIBLIOGRAPHY}}\\
\bibitem{1} H.~Georgi, \textit{Lie algebras in particle physics. From isospin to unified theories}, (1982).
\bibitem{2} P.~Wiegmann, \textit{Lecture notes in Quantum Field Theory}, (University of Chicago, Chicago, 2016).
\bibitem{3} S.~Carroll, \textit{Spacetime and geometry: an introduction to general relativity}, (Pearson Education, Essex, 2013).
\bibitem{3} M.~Peskin,~D.~Schroeder \textit{An Introduction to quantum field theory}, (Addison-Wesley, Reading, USA, 1995).
\end{flushleft}

\end{document}